\documentclass[]{article}

\usepackage{amsmath}
\usepackage{latexsym}
\usepackage{amsfonts}
\usepackage{amssymb}
\usepackage{amsthm}
\usepackage{bbm,dsfont}
\usepackage{graphicx}

\newtheorem{proposition}{Proposition}
\newtheorem{theorem}{Theorem}
\newtheorem{lemma}{Lemma}
\newtheorem{corollary}{Corollary}

\theoremstyle{definition}

\newcommand{\R}{\mathbb R} 
\newcommand{\C}{\mathbb C} 

\newcommand{\hi}{\mathcal{H}} 
\newcommand{\sh}{\mathcal{S(H)}} 
\newcommand{\eh}{\mathcal{E(H)}} 
\newcommand{\ip}[2]{\left\langle\,#1\,|\,#2\,\right\rangle} 
\newcommand{\kb}[2]{|#1\,\rangle\langle\,#2|} 
\newcommand{\no}[1]{\left\|#1\right\|} 
\newcommand{\tr}{\textrm{tr}} 
\newcommand{\id}{\mathbbm{1}} 
\newcommand{\nul}{\mathbbm{O}}
\newcommand{\fii}{\varphi}

\newcommand{\va}{\mathbf{a}} 
\newcommand{\vb}{\mathbf{b}} 
\newcommand{\vt}{\mathbf{t}} 
\newcommand{\vsigma}{\boldsymbol{\sigma}} 

\newcommand{\A}{\mathcal{A}}
\newcommand{\B}{\mathcal{B}}
\newcommand{\E}{\mathcal{E}}
\newcommand{\F}{\mathcal{F}}

\newcommand{\D}{\mathfrak{D}}
\newcommand{\U}{\mathfrak{U}}
\renewcommand{\S}{\mathfrak{S}}
\newcommand{\W}{\mathfrak{W}}

\newcommand{\sa}{\sigma_A}
\newcommand{\sap}{\sigma_{A'}}
\newcommand{\saap}{\sigma_{AA'}}

\newcommand{\eps}{\epsilon}

\begin{document}

\title{On the Sharpness and Bias of Quantum Effects}


\author{Paul Busch\\
{\small Department of Mathematics, University of York, UK\thanks{Electronic 
address: pb516@york.ac.uk}}
}

\date{\small\the\day/\the\month/\the\year}

\maketitle

\begin{abstract}
\noindent
The question of quantifying the {\em sharp\-ness} (or unsharpness) of 
a quantum mechanical effect  is investigated. Apart from sharpness, another property,  
{\em bias}, is found to be relevant for the joint measurability or coexistence of two effects.
Measures of bias will be defined and examples given. 
\end{abstract}

\section*{Dedication}
The impossibility of measuring jointly certain pairs of observables is an intriguing
non-classical feature of quantum theory that Pekka Lahti identified as a candidate for
a rigorous formulation of the principle of complementarity. While he was investigating 
this fundamental no-go statement in the early 1980s, he learned from Peter Mittelstaedt that
one of his students was aiming to prove the positive possibility of approximate joint measurements
of complementary quantities such as position and momentum. Pekka joined our
group as an Alexander von Humboldt Fellow, and together we found that a reconciliation 
between complementarity and (approximate) joint measurability is possible on the basis of 
the generalized representation of observables as positive operator measures (POMs). 
Since then we have pursued together our aspirations of understanding quantum mechanics and
understanding Nature. I have benefited much from Pekka's intellectual rigor and have been 
privileged ever since to enjoy his warm humanity. It is a great pleasure to present this paper to 
Pekka as a token of thanks and friendship on the occasion of his 60th birthday, with all good wishes
for many happy recurrences and productive years to come.

\section{Introduction}
The general description of quantum observables as positive operator measures (POMs) 
gives rise to a host of new operational possibilities not available within the set of standard 
observables (represented as projection valued measures). Here we focus on a particular 
issue of foundational significance: the possibility of joint measurements 
of certain pairs of noncommuting observables. 

Two observables are considered to be jointly measurable if there is a third, {\em joint observable}, 
of which the given observables are marginals. According to a well known theorem of von Neumann
\cite{MGQ}, two observables represented as projection valued measures are jointly measurable
if and only if they are mutually commuting \cite{Ylinen85}. Among pairs of general observables, 
commutativity is still a sufficient but no longer necessary condition for joint measurability.

Observables represented as projection valued measures are commonly understood to 
correspond to measurements with perfect accuracy; hence they can be called {\em sharp} 
observables. This characterization renders all other observables {\em unsharp}. If at least 
one of a pair of observables is sharp, then joint measurability cannot hold unless the two 
observables commute \cite{KrMu87}. It follows that for two {\em noncommuting} observables 
to be jointly measurable, it is necessary that {\em both} of them are unsharp.

The problem thus arises of determining the factors that are relevant for the 
characterization of jointly measurable pairs of noncommuting observables. In light of the above 
general observation one would expect a trade-off to hold between the {\em degrees} of 
noncommutativity and sharpness within the set of jointly measurable pairs of observables. 
This leads to the task of defining appropriate measures of (un)sharpness. We will focus on 
the case of {\em simple observables}, that is, POMs generated by a resolution of the identity 
$\id$ of the form $\{A,\id- A\}$, where both $A$ and $\id-A$ are positive operators, also referred to 
as {\em effects}.

We will consider candidates of sharpness measures in 2-dimensional Hilbert spaces, to begin with. 
A recently found \cite{BuSchm08,StReHe08,Liu-etal08} criterion of the joint measurability of pairs of  
qubit effects is seen to involve a trade-off between the degrees of noncommutativity, sharpness and 
yet another quantity called bias. The task of extending measures of sharpness and bias to arbitrary 
Hilbert spaces is not entirely trivial due to ambiguity and the fact that the 2-dimensional case is too 
simple to reveal relevant features. Nevertheless we have been able to identify several distinct 
measures of sharpness and bias, applicable in Hilbert spaces of arbitrary dimensions. 

\section{Preliminaries}

Our investigation is based on the usual quantum mechanical description
of a physical system represented by a complex separable Hilbert space $\hi$ with 
inner product $\ip{\fii}{\psi}$, $\fii,\psi\in\hi$. States are represented by positive 
operators $T$ of trace equal to one, the convex set of all states being denoted
$\sh$. The extremal elements of $\sh$ are the vector states, that is
the rank-one projections $T=\kb{\fii}{\fii}$, where $\fii$ is any
unit vector of $\hi$. An {\em effect} is a
selfadjoint operator $A$ on $\hi$ satisfying 
$\nul\le A\le \id$. Here $\nul$ and $\id$ are the null and identity
operators, respectively, and the partial order $A\le B$ is defined
as $\ip{\fii}{A\fii}\le\ip{\fii}{B\fii}$ for all
$\fii\in\hi$. An effect $A$ together with a state $T$
gives the number $\tr[T A]\in[0,1]$, which is the
probability for a measurement outcome represented by $A$ to occur
in a measurement performed on the state $T$.

The set of effects $\eh$ is thus the operator interval $[\nul,\id]$
with respect to $\le$.  $\eh$ is a convex set and its
extremal elements are exactly the (orthogonal) projections ($A=A^2$).
$\eh$ contains the convex subset of  {\em trivial} effects $\lambda\id$, $\lambda\in[0,1]$.

Projections will be called {\em sharp}, or {\em crisp effects}; all
other effects are called {\em unsharp} or {\em fuzzy}. The unsharp trivial effects
represent the extreme case of unsharpness;  their expectation values 
provide no information about the state of a system.
We note that an effect is a nontrivial sharp effect if and only if its spectrum consists of the two 
maximally separated elements 0,1; an effect is trivial if and only if its spectrum is a singleton.
The intersection of the sets of sharp and trivial effects is $\{\nul,\id\}$.

The {\em complement} of an  effect $A$ is defined as $A':=\id-A$. 
An algebraic relation that distinguishes sharp and unsharp effects arises from
the inequality $A^2\le A$ which characterizes the effects
among the selfadjoint operators. Hence we have $AA'\ge \nul$ exactly when $A$ is an 
effect, and an effect $A$ is sharp exactly when $AA'=\nul$.

The product of $A$ and $A'$ can be written as
$AA'=A^{1/2}A'A^{1/2}$, and this suggests the following operational
interpretation of the sharpness or unsharpness of an effect. Let
$\phi_L^A$ be the L\"uders operation associated with the effect $A$,
corresponding to an ideal measurement  (see \cite{QTM}) of the simple
observable given by $A,A'$, that is, $\phi_L^A(T)=A^{1/2}T
A^{1/2}$ for any state $T$. The probability of a measurement
giving an outcome associated with the effect $A$ is given by
$\tr[TA]=\tr[\phi_L^A(T)]$, and the sequential joint
probability  $\mathfrak{p}_T(A,A')$ that a L\"uders measurement of a simple observable $\A$
repeated in immediate succession yields first the outcome associated
with $A$ and then the outcome associated with $A'$ is
\begin{equation}
\mathfrak{p}_T(A,A')=\tr\left[\phi_L^A(T)A'\right]=\tr\big[T\,A^{1/2}A'A^{1/2}\big]=\tr[T\,AA'].
\end{equation}
This joint probability is zero for all states if and only if $A$ is a sharp effect. In this case, 
the measurement is repeatable for all states.  For an unsharp effect
the above joint probability is positive in some states. 
The trivial effects are characterized as those effects for which the above
joint probability is state independent. We note that the sets of sharp and trivial effects
have just the trivial projections $\nul,\id$ in common. For all other trivial effects $A=a_0\id$,
the joint probability $\mathfrak{p}_T(A,A')=a_0(1-a_0)$ is nonzero.


\section{Measures of sharpness and bias}

\subsection{Defining properties of a sharpness measure}

We seek a definition of a measure of the {\em sharpness}
of an effect $A$ (and its associated simple observable $\A$) which singles out and disinguishes 
the trivial effects on the one hand and the nontrivial sharp ones on the other
hand. Specifically a function $\eh\ni A\mapsto \S(A)$ will be accepted as a sharpness 
measure if it satisfies (at least) the following requirements:\\
(S1) $0\le \S(A)\le 1$;\\
(S2) $\S(A)=0$ if and only if $A$ is a trivial effect;\\
(S3) $\S(A)=1$ if and only if $A$ is a nontrivial projection;\\
(S4) $\S(A')=\S(A)$;\\
(S5) $\S(CAC^{-1})=\S(A)$ for all invertible operators $C$;\\
(S6) $A\mapsto \S(A)$ is (norm) continuous.\\
The function $A\mapsto \U(A)=1-\S(A)$ can then be taken as a measure of unsharpness or fuzziness. 
It satisfies a similar set of conditions, with appropriate adjustments.

Condition (S1) is merely a convention. (S2) and (S3) are the decisive properties of any 
sharpness measure once (S1) is adopted. (S4) and (S5) are motivated by the idea that 
the sharpness of an effect $A$ 
is determined by the distribution of the spectrum $\sa$ of $A$. (S6) reflects the idea 
that a small change of $A$ results in a small change of the degree of sharpness. 
(S5) and (S6) are automatically satisfied for sharpness measures $\S(A)$ defined 
as functions of expressions such as $\no{A}$, $\no{A'}$, $\no{AA'}$.

We will see that the above list does not single out a unique sharpness measure.
One could thus conceive of further properties that a sharpness measure may be
required to have. For example, it could be argued that a convex combination of two effects
cannot be sharper than these two effects themselves; this would be true if
$A\mapsto \S(A)$ is an affine functional. In view of the 
operational interpretation of unsharpness considered in the preceding section, we may have to be prepared to take into account that a measure $\S(A)$ will depend on properties of $AA'$.

\subsection{Bias of an effect}

A yes-no experiment would be regarded as {\em biased} if one of the outcomes turned out to be preferred,
whatever the preparation. We may thus define an effect $A$ to be {\em unbiased} if for every state $T$ there
is a state $T'$ such that $\tr{TA}=\tr{T'A'}$.  This is equivalent to saying that the convex hulls of $\sa$ and $\sap$ are
identical, or that the midpoints of $\sa$ and $\sap$ are the same. For later use we define the minimum, maximum, width, and 
midpoint of $\sa$, respectively as:
\begin{eqnarray}
M(\sa)&:=&\max(\sa)=\|A\|;\\
m(\sa)&:=&\min(\sa)=1-\max(\sap)=1-\|A'\|;\\
\W(\sa)&:=&M(\sa)-m(\sa)=\|A\|+\|A'\|-1;\\
\mu(\sa)&:=&\tfrac 12\left(M(\sa)+m(\sa)\right)=\tfrac12\left(\|A\|+1-\|A'\|\right).
\end{eqnarray}
Then $A$ is unbiased according to the above condition if $\mu(\sa)=\frac 12$. 
We note that the set of unbiased effects is a convex subset of $\eh$.
The {\em maximally biased} effects are $\nul$ and $\id$.

A measure of bias $\eh\ni A\mapsto \B(A)$will be understood as a function that satisfies (at least)
the following conditions:\\
(B1) $-1\le\B(A)\le 1$;\\
(B2) $\B(A)=0$ if and only if $A$ is unbiased, i.e., $\mu(\sa)=\frac 12$;\\
(B3) $\B(A)=1$ if and only if $A=\id$, and $\B(A)=-1$ if and only if $A=\nul$;\\
(B4) $\B(A)=-\B(A')$;\\
(B5) $\B(CAC^{-1})=\B(A)$ for all invertible operators $C$;\\
(B6) $A\mapsto \B(A)$ is (norm) continuous.\\
The motivation for these postulates is similar to those for sharpness measures. The last two can be 
secured by defining $\B(A)$ as a function of quantities such as $\no{A}$, $\no{A'}$, $\no{AA'}$.

\section{Sharpness and bias measures for qubit effects}

\subsection{Examples of sharpness measures}

We first construct examples of sharpness measures for qubit effects.

A qubit is described in a 2-dimensional Hilbert space, $\hi=\C^2$. We use the Pauli basis
$\{\id,\sigma_1,\sigma_2,\sigma_3\}$ to represent an operator $A$ as $A=a_0\id+\va\cdot\vsigma$.
Operator $A$ is selfadjoint if and only if $(a_0,\va)\in\R^4$. A state operator $T$ is given by
$T=\frac 12(\id+\vt\cdot\sigma)$, where the Euclidean norm  of $\vt$ satisfies $0\le|\vt|\le 1$.
$A$ is an effect if and only if its eigenvalues are between 0 and 1, that is, $0\le a_0\pm|\va|\le 1$.
An effect $A$ is a projection if and only if $a_0=|\va|=\frac 12$.

We recall that an effect $A$ is nontrivial and sharp exactly when its spectrum $\sa=\{0,1\}$, and $A$ is trivial
exactly when $\sa=\{a_0\}$. We seek a function $\E(\C^2)\ni A\mapsto \S(A)\in[0,1]$ which assumes value 1 exactly in the former case and value 0 exactly in the latter case. Such a function is easily obtained: we simply take its value to be the {\em spectral width}, that is, the difference between the largest and smallest eigenvalue:
\begin{equation}
\S_a^{(2)}(A):=\W(\sa)=(a_0+|\va|)-(a_0-|\va|)=2|\va|.
\end{equation}
The superscript indicates that this quantity is specific to the case of a 2-dimensional Hilbert space. 
It is easily verified that $\S_a^{(2)}$ satisfies all conditions (S1)-(S6); in addition, it is convex.

This function extends in an obvious way to higher dimensions:
\begin{equation}
\S_a(A):=\W(\sa)=\|A\|-(1-\|A'\|)=\|A\|+\|A'\|-1.
\end{equation}
In $\hi=\C^2$ we have $\S_a=\S_a^{(2)}$, and this function is a sharpness measure. However,
in higher dimensions $\S_a$ fails to satisfy (S3): if $A$ is an effect with spectrum $\{0,\alpha,1\}$, with
$0<\alpha<1$, then $\S_a(A)=1$ although $A$ is not sharp.

Another simple function that constitutes a sharpness measure in $\C^2$ is the following:
\begin{equation}
\S_b^{(2)}(A):=4\min(a_0,1-a_0)\,|\va|=2|\va|\,[1-|2a_0-1|].
\end{equation}
We note that $A=a_0\id+\va\cdot\vsigma$ is an effect if and only if 
\begin{equation}\label{eqn:effect}
0\le|\va|\le\min(a_0,1-a_0).
\end{equation} 
Putting $\xi:=\min(a_0,1-a_0)$, $\eta:=|\va|$, it is straightforward to see that the 
function $(\xi,\eta)\mapsto 4\xi\eta$, defined on the domain 
\begin{equation}\label{eqn:D}
D:=\{(\xi,\eta)\,:\, 0\le \xi\le \frac 12,\ 0\le \eta\le \xi\},
\end{equation}
assumes its minimum 0 exactly 
on the line segment in $D$ given by $\eta=0$, while the maximum 1 is reached 
exactly at the point $(\xi,\eta)=(\frac 12,\frac 12)$. This proves the properties 
(S1), (S2) and (S3). The symmetry (S4) is given by construction. 
Since 
\begin{equation}\label{eqn:a0-va}
a_0=\tfrac 12(a_0+|\va|)+\tfrac 12(a_0-|\va|)=\mu(\sa),\quad
|\va|=\tfrac 12(a_0+|\va|)-\tfrac 12(a_0-|\va|)=\tfrac 12\W(\sa),
\end{equation}
it is seen that $\S_b^{(2)}$ depends only on the eigenvalues of $A$ and thus 
(S5) is fulfilled. For the same reason, $\S_2^{(2)}$ is norm continuous.

We can characterize a sharpness measure $\S^{(2)}$ in $\C^2$ more 
systematically as follows. $\S^{(2)}(A)$ should depend on the eigenvalues 
of $A$ (in fulfillment of (S5)) and thus can be expressed as a function 
$f(a_0,|\va|)$ in view of (\ref{eqn:a0-va}). The domain of $f$ is given by 
Eq.~(\ref{eqn:effect}). Next, the condition (S4) reads $f(a_0,|\va|)=f(1-a_0,|\va|)$, 
which entails that $\S^{(2)}$ can in fact be written as a function of 
$\xi=\min(a_0,1-a_0)$ and $\eta=|\va|$, so that $\S^{(2)}(A)=f(\xi,\eta)$, with the domain 
now being $D$ as given in (\ref{eqn:D}). Any continuous function $f$ on $D$
with the property that the maximum 1 is assumed exactly on $(\frac 12,\frac 12)$ and
the minimum 0 is assumed exactly on the points $(\xi,0)$ gives rise to a sharpness measure.

Returning to the example of $\S^{(2)}_b$, we note that this can be written in
a form that lends itself to generalization to arbitrary Hilbert spaces:
\begin{equation}\label{eqn:sb}\begin{split}
\S_b(A)&=4\min\left[\tfrac 12(\|A\|+1-\|A'\|),1-\tfrac 12(\|A\|+1-\|A'\|)\right]\,
\tfrac 12(\|A\|+\|A'\|-1)\\
&=\min\left[1+\|A\|-\|A'\|,1- (\|A\|-\|A'\|)\right]\,(\|A\|+\|A'\|-1)\\
&=2\min\bigl(\mu(\sa),\mu(\sap)\bigr)\,\W(\sa)\\
&=\left(1-\left|\,\|A\|-\|A'\|\,\right| \right)\,(\|A\|+\|A'\|-1).
\end{split}
\end{equation}
As with $\S_a$, we find again that $\S_b(A)=1$ if only $\|A\|=\|A'\|=1$, that is, if
$\{0,1\}\subseteq\sa$. So (S2) is violated if $\hi$ is at least 3-dimensional.
We thus find that it seems less than straightforward to find a sharpness 
measure for Hilbert spaces of arbitrary dimensions by extending a measure 
suitable for $\C^2$. 

Moreover,  a function like $\S_b^{(2)}$ will not even have,
in general, a unique extension. In fact, another extension of this function is
given by the following:
\begin{equation}\label{eqn:s1}
\S_1(A):=\|A\|+\|A'\|-\left[\|AA'\|+\|\id-AA'\|\right].
\end{equation}
The fact that in $\C^2$ we have $\S_1=\S_b^{(2)}$ is easily verified by noting that 
\begin{equation}
\|AA'\|=\tfrac 14-\min\left((\|A\|-\tfrac 12)^2,(\|A'\|-\tfrac 12)^2\right),\quad
\|\id-AA'\|=\tfrac 34+\max\left((\|A\|-\tfrac 12)^2,(\|A'\|-\tfrac 12)^2\right)
\end{equation}
We will show below that the function $\S_1$ is in fact a sharpness measure.

\subsection{Sharpness and bias measures from qubit coexistence}

Another sharpness measure was found in connection with a criterion for the
joint measurability for two qubit effects. Two effects $A,B$ are jointly measurable, or
coexistent, if there is a POM,  called a joint observable for $A,B$, whose range 
contains $A$ and $B$; this ensures that both resolutions of identity 
$\{A,A'\}$ and $\{B,B'\}$ are contained in the range of the joint observable.
This joint observable can always be taken to be generated by a resolution of the
identity of the form $\{G_{11},G_{12}G_{21},G_{22}\}$ so that
\[
A=G_{11}+G_{12},\quad B=G_{11}+G_{21}.
\]
It was recently proven independently by three groups that effects 
$A=a_0\id+\va\cdot\vsigma$ and $B=b_0\id+\vb\cdot\vsigma$ are coexistent
exactly when a certain inequality holds \cite{BuSchm08,StReHe08,Liu-etal08}. This 
inequality can be cast in the form \cite{BuSchm08}
\begin{equation}
\tfrac 12[\F(2-\B)+\B(2-\F)]+(xy-4\va\cdot\vb)^2\ge 1.
\end{equation}
Here the following abbreviations are used:
\begin{eqnarray}
\F&:=&\F^{(2)}(A)^2+\F^{(2)}(B)^2;\\
 \B&:=&\B^{(2)}(A)^2+\B^{(2)}(B)^2;\\
x&:=&\F^{(2)}(A)\B^{(2)}(A)=2a_0-1\equiv\B_a^{(2)}(A);\label{eqn:x}\\
y&:=&\F^{(2)}(B)\B^{(2)}(B)=2b_0-1\equiv\B_a^{(2)}(B);\\
\F^{(2)}(A)&:=&\sqrt{a_0^2-|\va|^2}+\sqrt{(1-a_0)^2-|\va|^2};\label{eqn:fa2}\\
\B^{(2)}(A)&:=&\sqrt{a_0^2-|\va|^2}-\sqrt{(1-a_0)^2-|\va|^2}.\label{eqn:ba2}
\end{eqnarray}
$\F^{(2)}(B)$ and $\B^{(2)}(A)$ are defined similarly.
It has been shown \cite{StReHe08,BuSchm08} that the following is a sharpness measure:
\begin{equation}\label{eqn:sc2}
\begin{split}
\S_c^{(2)}(A):=1-\F^{(2)}(A)^2&=2\left[a_0(1-a_0)+|\va|^2-\sqrt{(a_0^2-|\va|^2)((1-a_0)^2-|\va|^2)}\right]\\
&=2\left[a_0(1-a_0)+|\va|^2\right]-\sqrt{4\left[a_0(1-a_0)+|\va|^2\right]^2-4|\va|^2}.
\end{split}
\end{equation}
This function is even more complicated than the preceding ones. Still we will give 
a reformulation of this quantity so as to render it a sharpness measure for arbitrary 
Hilbert spaces.

It is also easy to see that $\B^{(2)}(A)$ and $\B_a^{(2)}(A)$ are both bias measures in $\C^2$.

\section{Sharpness and bias measures for arbitrary Hilbert spaces}

We begin with a simple spectral characterization of an effect which arises as a generalization of inequality (\ref{eqn:effect}).
\begin{proposition}
A selfadjoint bounded linear operator $A$ in $\hi$ is an effect if and only of the following inequality holds:
\begin{equation}\label{eqn:effect-gen}
0\le \|A\|+\|\id-A\|-1\le 1-\bigl|\,\|A\|-\|\id-A\|\,\bigr|\,.
\end{equation}
\end{proposition}
\noindent Note that for an effect $A$ this relation can be written as (cf.~Eq.~\ref{eqn:sb})
\begin{equation}\label{eqn:effect-gen2}
0\le\tfrac 12\W(\sa)\le \min\left(\mu(\sa),\mu(\sap)\right).
\end{equation}
This shows that the statement is indeed quite obvious: this inequality ensures that the spectrum of $A$ lies in the 
interval $[0,1]$. 

We proceed to construct some relatively simple measures of sharpness and bias for general Hilbert spaces. 

\begin{theorem}
The function $\S_0$ given by
\begin{equation}\label{eqn:s0}
\S_0(A):=\|A\|\,\|A'\|-\|AA'\|
\end{equation} 
is a sharpness measure.
\end{theorem}
\begin{proof}
It is obvious that $\S_0(A)\in[0,1]$.\\
We write
\begin{equation}
M:=\max\sa=\|A\|,\quad m:=\min\sa=1-\|A'\|.
\end{equation}
Now, if $\S_0(A)=0$, we have $\|A\|\,\|A'\|=\|AA'\|$. This reads:
\[
M(1-m)=\max_{\lambda\in\sa}\lambda(1-\lambda).
\]
But here $\lambda\le M$ and $1-\lambda\le1-m$, thus the above equation requires that
$\lambda=M$ and $\lambda=m$, hence $m=M$ and $A=m\id$. This proves (S2).\\
Next, suppose $\S(A)=1$. This entails $\|A\|\,\|A'\|=1$ and $\|AA'\|=0$, so that $A$ is
a nontrivial projection. Thus (S3) is verified.
\end{proof}

\begin{theorem}
The function 
\begin{equation}
\eh\ni A\mapsto \B_0(A):=2\mu(\sa)-1=\|A\|-\|A'\|
\end{equation}
is a bias measure.
\end{theorem}
\noindent The proof is trivial and will be omitted.

We can now give an equivalent way of phrasing the inequality (\ref{eqn:effect-gen2}) characterizing effects.
\begin{corollary}
A bounded selfadjoint operator $A$ is an effect if and only if
\begin{equation}
\W(\sa)+\bigl|\,\B_0(A)\,\bigr|\le 1.
\end{equation}
\end{corollary}
\noindent Effects are thus distinguished by a trade-off between spectral width and bias.

In order to introduce the next sharpness measure, we note two relevant features of the spectrum 
$\sa$ of $A$. The first is the {\em spectral width}, which we denoted $\W(\sa)$ and defined as the length 
of the smallest interval containing $\sa$. The second feature is the extent to which
$A$ deviates from the extreme cases of projections
($\sa\subseteq\{0,1\}$) or trivial effects ($\sa=\{\lambda_0\}$).
This latter feature will be called the {\em (spectral) dispersion} $\D(\sa)$
of $A$ or of $\sa\subseteq [0,1]$; from the above discussion
of the significance of the operator $AA'$, it is to be expected that $\D(\sa)$ 
is related to the spectral width of $AA'$. Thus we define:
\begin{equation}
\D(\sa):=\W(\saap).\label{eqn:disp1}
\end{equation}
To analyze these concepts, we recollect some simple observations.
\begin{eqnarray}
\W(\sa)&=&\no{A}+\no{A'}-1=\W(\sap);\label{eqn:width2}\\
\D(\sa)&=&\no{AA'}+\no{\id-AA'}-1=\D(\sap).\label{eqn:disp2}
\end{eqnarray}
We thus have $\S_a(A)=\W(A)$ and $\S_1(A)=\W(\sa)-\D(\sa)$.
\begin{lemma}\label{lem:width}
Let $A$ be an effect. Then:
\begin{equation}
\W(\sa)=2\min_{\kappa\in\R}\no{A-\kappa\id}.
\end{equation}
The minimum is assumed at $\kappa=a_0$, where
\begin{equation}
a_0=\tfrac 12\left[\|A\|-\|A'\|+1\right]=\mu(\sa).
\end{equation}
\end{lemma}
\begin{proof}
We calculate $ \no{A-\kappa\id}=\sup\{|\ip{\fii}{A\fii}-\kappa|:\fii\in\hi,\|\fii\|=1\}$ for the various possible cases.\\
(i) $\kappa\ge\no{A}$:
\[
\no{A-\kappa\id}=\kappa-\min\sa\ge\no{A}-[1-\no{A'}]\quad(\mathrm{attained\ at\ }\kappa=\no{A});\
\]
(ii) $0\le\no{A}-\kappa\le\kappa-[1-\no{A'}]$:
\[
\no{A-\kappa\id}=\kappa-\min\sa\ge\tfrac 12 [\no{A}-[1-\no{A'}]\quad
 (\mathrm{attained\ at\ } \kappa=  \tfrac 12 [\no{A}+[1-\no{A'}]);
 \]
 (iii) $ \no{A}-\kappa\ge\kappa-[1-\no{A'}]\ge 0$:
 \[
\no{A-\kappa\id}=\no{A}-\kappa\ge \tfrac12 [\no{A}-[1-\no{A'}]\quad
 ( \mathrm{attained\ at\ }\kappa=  \tfrac 12 [\no{A}+[1-\no{A'}]);
 \]
 (iv) $ [1-\no{A'}]\ge \kappa$:
 \[
 \no{A-\kappa\id}=\no{A}-\kappa \ge \no{A}-[1-\no{A'}]\quad
 (\mathrm{attained\ at\ }\kappa=[1-\no{A'}]).
 \]
From this it is seen that $\min_\kappa\no{A-\kappa\id}=\frac
12(\no{A}+\no{A'}-1)=\frac 12\W(\sa)$.\\
Using the value $\kappa=a_0$, we find
\begin{equation*}
\|A-a_0\id\|=\max_{\lambda\in\sa}|\lambda-a_0|=
\|A\|-a_0=\tfrac 12\left[\|A\|+\|A'\|-1\right].
\end{equation*}
\end{proof}
This shows that the spectral width of an effect $A$ is twice the
minimal norm distance of $A$ from the set of trivial effects.
\begin{proposition}
Let $A$ be an effect; then:\\
(a) $0\le\W(\sa)\le 1$;\\
(b) $\W(\sa)=1$ iff $\{0,1\}\subseteq\sa$;\\
(c) $\W(\sa)=0$ iff $A$ is trivial.
\end{proposition}
\begin{proof}
Inequality (a) is an immediate consequence of the definition of $\W(\sa)$.\\
(b): For $\W(\sa)$ to be equal to 1 it is necessary and sufficient
that $\max\sa=1$ and $\min\sa=0$.\\
(c): $\W(\sa)=0$ is equivalent to $\max\sa=\min\sa$, which means that
$\sa$ is a one-point set or that $A$ is a constant operator.
\end{proof}

\begin{lemma}\label{lem:prop-of-AA'}
For an effect $A$ the operator $AA'$ has the following properties:\\
(a) $\nul\le AA'\le\frac 14\id$;\\
(b) $\frac 14-\max\big\{(\no{A}-\tfrac 12)^2,
 (\no{A'}-\tfrac 12)^2\big\}\le \no{AA'}\le\frac 14$;\\
(c) $\no{AA'}=\frac 14$ if and only if $\frac 12\in\sa$;\\
(d) $\no{\id-AA'}= \frac 34+
 \max\big\{(\no{A}-\tfrac 12)^2,(\no{A'}-\tfrac 12)^2\big\}$.
\end{lemma}
\begin{proof}
Let $\lambda\mapsto E^A_\lambda$ denote the spectral family of $A$. Then
\begin{equation}\label{eqn:spec-AAdash}
AA'=\int\lambda(1-\lambda)dE^A_\lambda=\tfrac14\id-\int(\lambda-\tfrac
12)^2dE^A_\lambda=\tfrac 14\id-(A-\tfrac 12\id)^2.
\end{equation}
Using this and the fact that $\lambda(1-\lambda)\le \frac 14$, with
the maximum attained at $\lambda=\frac 12$, we obtain
immediately (a),(b)  and the sufficiency part of $(c)$. If $\frac
12\not\in\sa$ then there is a $\lambda_0\in\sa$ which assumes the
shortest positive distance of the closed set $\sa$ from $\frac 12$;
it follows that $\no{AA'}=\lambda_0(1-\lambda_0)<\frac 14$. This
proves the necessity part of (c). To show (d), we note that
$\lambda\mapsto\lambda(1-\lambda)$ is increasing for
$0\le\lambda\le\frac 12$ and decreasing for $\frac 12\le\lambda\le
1$; then $\min\{\lambda(1-\lambda):\lambda\in\sa\}$ is attained
either at $\lambda=\min\sa$ or $\lambda=\max\sa$, hence:
\begin{eqnarray*}
\no{\id-AA'}
&=&\max\{1-\no{A}(1-\no{A}),1-\no{A'}(1-\no{A'})\}\\
&=&\tfrac 34+\max\left\{\left(\no{A}-\tfrac 12\right)^2,\left(\no{A'}-\tfrac 12\right)^2\right\}.
\end{eqnarray*}
\end{proof}
\begin{proposition}
Let $A$ be an effect. Then:\\
(a) $\D(\sa)$ is given by
\begin{equation}\label{eqn:D-explicit}
\D(\sa)=\max\left\{\left(\no{A}-\tfrac 12\right)^2,\left(\no{A'}-\tfrac 12\right)^2\right\}
-(\tfrac 14-\no{AA'});
\end{equation}
(b) $0\le\D(\sa)\le\frac 14$;\\
(c) $\D(\sa)=\frac 14$ iff $\frac 12\in\sa$ and $1\in\sa$ or $0\in\sa$.\\
(d) $\D(\sa)=0$ iff $\sa=\{\lambda\}$ or $\sa=\{\lambda,1-\lambda\}$
(with $0\le\lambda\le 1$).
\end{proposition}
\begin{proof}
The first equation is a direct consequence of Lemma \ref{lem:prop-of-AA'}.\\
It follows directly from the definition that $\D(\sa)\ge0$. The
maximum of $\D(\sa)$ is obtained by maximizing the positive term in
Eq.~(\ref{eqn:D-explicit}) and minimizing the negative term: this
happens when either $\no{A}=1$ or $\no{A'}=1$, and $\frac 12\in\sa$.
This yields $\max_{A\in\eh}\D(\sa)=\frac 14$.

Next, $\D(\sa)=0$ means that $AA'$ is a trivial effect (Lemma \ref{lem:width}). Using again
the spectral representation (\ref{eqn:spec-AAdash}) of $AA'$, we see that $AA'$ is
a multiple of $\id$ if and only if $\sa$ is such that
$\lambda(1-\lambda)$ is a constant $\frac 14-\eps^2$. This is the
case if and only if either $\sa=\{\lambda\}$ or
$\sa=\{\lambda,1-\lambda\}$ (where $0\le\lambda\le 1$).
\end{proof}

This result shows that $\D(\sa)$ becomes large only if $\sa$ extends to at least
one of the end points of the interval [0,1] {\em and} to its center.

\begin{theorem}
For any effect $A$ the following holds true:\\
(a) $0\le\W(\sa)-\D(\sa)\le1$;\\
(b) $\W(\sa)-\D(\sa)=0$  if and only if $A$ is trivial;\\
(c) $\W(\sa)-\D(\sa)=1$ if and only if $A$ is a nontrivial projection.
\end{theorem}
\begin{proof}
We know already that $\W(\sa)\le 1$ and $\D(\sa)\ge 0$; this gives $\W(\sa)-\D(\sa)\le 1$.\\
To prove that this quantity is nonnegative we use the explicit form:
\begin{equation}\begin{split}
\W(\sa)-\D(\sa)=&\no{A}-\tfrac 12+\no{A'}-\tfrac12+\left[\tfrac 14-\no{AA'}\right]\\
&\quad-\max\{(\no{A}-\tfrac 12)^2,(\no{A'}-\tfrac 12)^2\}.
\end{split}
\end{equation}
We consider the following cases.\\
(i) $\epsilon\equiv\no{A}-\tfrac 12\ge 0$, $\epsilon'\equiv\no{A'}-\tfrac 12\ge 0$.\\
This case entails that $\sa\subseteq[\frac 12-\eps',\frac 12+\eps]$,
$\sap\subseteq[\frac 12-\eps,\frac 12+\eps']$, where the interval
boundaries are the minima and maxima of the spectra.
Then using $\no{AA'}\le\frac 14$ we estimate:
\begin{eqnarray*}
\W(\sa)-\D(\sa)&=& \epsilon+\epsilon'-\max\{\eps^2,\eps'^2\}+\left[\tfrac14-\no{AA'}\right]\\
&\ge&(\eps+\eps')-\tfrac 12[\eps^2+\eps'^2+|\eps^2-\eps'^2|]\\
&=&\left\{\begin{matrix}\eps(1-\eps)+\eps'&\text{if\ }\eps\ge\eps';\cr
\eps'(1-\eps')+\eps&\text{if\ }\eps\le\eps'.\end{matrix}\right.
\end{eqnarray*}
This becomes equal to 0 exactly for $\eps=\eps'=0$, that is, $A=A'=\frac 12\id$.\\
(ii) $-\tfrac12 \le-\eps\equiv\no{A}-\tfrac 12\le 0\le \eps'\equiv\no{A'}-\tfrac 12\le\tfrac 12$.\\
This condition entails that $\sa\subseteq [\tfrac 12-\eps',\tfrac 12-\eps]$ and
$\sap\subseteq[\tfrac 12+\eps,\tfrac 12+\eps']$. Here the bounds are the minima
and maxima of the spectra. Furthermore, $\no{AA'}=\no{\int\lambda(1-\lambda)dE^A_\lambda}=
\frac 14-\eps^2$.
Then $\W(\sa)=\eps'-\eps \ge 0$, and $\D(\sa)=
\max\{\eps^2,\eps'^2\}-(\tfrac 14-\no{AA'})=\eps'^2-\eps^2=(\eps'-\eps)(\eps'+\eps)$. Thus,
\begin{equation*}
\W(\sa)-\D(\sa)=(\eps'-\eps)[1-(\eps+\eps')]\ge 0.
\end{equation*}
The latter expression is  0 exactly when $\eps'=\eps$, that is, when $\sa=\{\frac 12-\eps\}$, 
i.e., when $A$ is trivial.\\
(iii) $-\tfrac12 \le-\eps':=\no{A'}-\tfrac 12\le 0\le \eps'=\no{A}-\tfrac 12\le\tfrac 12$.\\
This case is analogous to the previous one, with $A$ and $A'$ exchanged.\\
(iv) $\no{A}< \frac12$ and $\no{A'}<\frac 12$.\\
This case does not arise since $\no{A}<\frac 12$ implies $\no{A'}>\frac 12$.

Finally, assume that $\W(\sa)-\D(\sa)=1$. This is equivalent to
$\W(\sa)=1$ and $\D(\sa)=0$. From our previous results this is
equivalent to $\{0,1\}\subseteq\sa$ and $A$ being either trivial or
having $\sa=\{\frac 12\pm\eps\}$. The first condition rules out
triviality and requires $\eps=0$. It follows that $A$ must be a
nontrivial projection. Conversely, it is obvious that this condition
entails that $\W(\sa)=1$ and $\D(\sa)=0$.
\end{proof}

The following is now immediate.
\begin{corollary}
The function $\eh\ni A\mapsto\S_1(A)$, defined by
\begin{equation}
\S_1(A):=\W(\sa)-\D(\sa)=\W(\sa)-\W(\saap),
\end{equation}
satisfies requirements (S1)-(S6). 
\end{corollary}
This proves that the extension $\S_1$ of $\S_b^{(2)}$ is a sharpness measure
for Hilbert spaces of arbitrary dimension.

\section{Further measures of sharpness and bias from the coexistence condition}

There are various extension of the sharpness measure $\S_c^{(2)}$. The first, $\S_2$, arises
from the observation that in $\C^2$ the following equations hold:
\begin{equation*}\begin{split}
2\|A\|\,\|A'\|-\W(\sa)&=2[a_0(1-a_0)+|\va|^2];\\
2\S_0(A)-\S_1(A)&=4|\va|^2.
\end{split}
\end{equation*}
This leads us to define:
\begin{equation}\label{eqn:s2}
\S_2(A):=2\|A\|\,\|A'\|-\W(\sa)-\sqrt{\left[2\|A\|\,\|A'\|-\W(\sa)\right]^2-\left[2\S_0(A)-\S_1(A)\right]}.
\end{equation}
\begin{theorem}\label{thm:s2}
The function $A\mapsto\S_2(A)$ is a sharpness measure.
\end{theorem}
\begin{proof}
It suffices to prove (S1)-(S3). Write $\S_2(A)$ in the form $\S_2(A)=X-\sqrt{X^2-Y}$.
We first show that 
\[
0\le X\equiv 2\|A\|\,\|A'\|-\W(\sa)\le 1,\quad 0\le Y\equiv2\S_0(A)-\S_1(A)\le X^2.
\]
This ensures that $\S_2(A)\in[0,1]$.
We observe:
\[
X=2\|A\|\,\|A'\|-\|A\|+(1-\|A'\|)=2M(1-m)-M+m=M+m-Mm=M(1-m)+m(1-M)\in[0,1].
\]
Next we compute:
\begin{equation*}\begin{split}
Y&=2\S_0(A)-\S_1(A)=2\|A\|\,\|A'\|-2\|AA'\|-\W(\sa)+\W(\saap)\\
&=(2\|A\|\,\|A'\|-\W(\sa))+(\W(\saap)-2\|AA'\|)\\
&=X+\|\id-AA'\|-1-\|AA'\|=X-2\mu(\saap)\\
&=M(1-m)+m(1-M)+\tfrac 34+\max\left\{(M-\tfrac 12)^2,(m-\tfrac 12)^2\right\}
-1-\tfrac 14+\min_{\lambda\in\sa}(\lambda-\tfrac 12)^2\\
&=M(1-m)+m(1-M)-\tfrac 12+\min_{\lambda\in\sa}(\lambda-\tfrac 12)^2
+\max\left\{(M-\tfrac 12)^2,(m-\tfrac 12)^2\right\}.
\end{split}
\end{equation*}
We consider the following cases for 
\[
\Delta\equiv(M-\tfrac 12)^2-(m-\tfrac 12)^2=(M-m)(M+m-1):
\]
(I) $\Delta=0$, i.e., $M=m$ or $M=1-m$;\\
(II) $\Delta>0$, i.e. $M>m$ and $M>1-m$;\\
(III) $\Delta<0$, i.e. $M>m$ and $M<1-m$.\\
Also put $\min_{\lambda\in\sa}(\lambda-\frac 12)^2\equiv\delta$.

\noindent Case (I) with $M=m$: then $\delta=(M-\frac 12)^2$ and 
$Y=2M(1-M)-\tfrac 12+2(M-\frac 12)^2=0$.

\noindent Case (I) with $M=1-m$: then $Y=3(\frac 14-M(1-M))+\delta\ge 0$.

\noindent Case (II): In this case we must have $M>\frac 12$ (since $m<M$ and $M+m>1$). Then one
obtains:
\[
Y=(M-2m+\tfrac 12)(M-\tfrac 12)+\delta =\left\{\begin{matrix}
 (M-m+\tfrac 12-m)(M-\tfrac 12)+\delta\ge 0&\text{if\ }m\le\tfrac 12;\\
 &\\
 (M-\tfrac 12-2(m-\tfrac 12))(M-\tfrac 12)+(m-\tfrac 12)^2=(M-m)^2&\text{if\ } m\ge\tfrac 12.
\end{matrix}\right.
\]

\noindent Case (III): In this case we must have $m<\tfrac 12$. One obtains:
\[
Y=(2M-m-\tfrac 12)(\tfrac 12-m)+\delta=\left\{\begin{matrix}
(M-\tfrac 12+M-m)(\tfrac 12-m)+\delta\ge 0&\text{if\ } M\ge\tfrac 12;\\
&\\
(\tfrac 12-m-2(\tfrac 12-M))(\tfrac 12-m)+(\tfrac 12-M)^2=(M-m)^2&\text{if\ }M\le\tfrac 12.
\end{matrix}\right.
\]
Thus we see that $Y\ge 0$ in all cases.

Next we show that $Y\le X^2$. Note that $X=2M(1-m)-M+m=M-2Mm+m$. Then
\begin{equation*}
X^2-Y=(M-2Mm+m)^2-Y.
\end{equation*}
We consider again the above cases, using expressions for $Y$ obtained there.\\
Case (I) with $M=m$: here $Y=0$, so $X^2-Y=X^2\ge 0$.\\
Case (I) with $M=1-m$: here we have
\begin{equation*}\begin{split}
X^2-Y&=(1-2M(1-M))^2-3(\tfrac 14-M(1-M))-\delta\\
&=4M^2(1-M)^2+(M-\tfrac 12)^2-\delta\ge4M^2(1-M)^2\ge0.
\end{split}
\end{equation*}
Case (II): Here we obtain:
\begin{equation*}\begin{split}
X^2-Y&=(M-2Mm+m)^2-[(M-2m+\tfrac 12)(M-\tfrac 12)+\delta]\\
&=(M-2Mm+m)^2-(M-m)^2+(m-\tfrac 12)^2-\delta\\
&=2M(1-m)2m(1-M)+(m-\tfrac 12)^2-\min_{\lambda\in\sa}(\lambda-\tfrac 12)^2\ge0.
\end{split}
\end{equation*}
Case (III): We have:
\begin{equation*}\begin{split}
X^2-Y&=(M-2Mm+m)^2-[(2M-m-\tfrac 12)(\tfrac 12-m)+\delta]\\
&=(M-2Mm+m)^2-(M-m)^2+(M-\tfrac 12)^2-\min_{\lambda\in\sa}(\lambda-\tfrac 12)^2\ge0.
\end{split}
\end{equation*}

Finally we check for which effects $A$ one has $\S_2(A)$ equal to 0 or 1. First consider $\S_2(A)=0$.
This implies that $Y=2\S_0(A)-\S_1(A)=0$. In case (I) this holds whenever $M=m$, and where $M=1-m$,
it requires $m=m=\tfrac 12$. It is also easily verified that in the cases (II) and (III) we always have $Y>0$.
Hence $\S_2(A)=0$ is equivalent to $A$ being trivial.

Assume now that $\S_2(A)=1$. This is equivalent to $X=1$ and $X^2-Y=0$. The first condition reads:
$M-2Mm+m=M(1-m)+m(1-M)=1$, which can only be satisfied by $M=1$, $m=0$. 
Hence $\sa\supseteq\{0,1\}$. This only leaves
case (I) with $M=1-m=1$ and thus $X^2-Y=4M^2(1-M)^2+(M-\tfrac 12)^2-\delta=\tfrac 14-\delta=0$. This 
finally entails $|\lambda-\tfrac 12|=\tfrac 12$ for all $\lambda\in\sa$ and so $\sa=\{0,1\}$, that is, $A$ is a 
nontrivial projection.
\end{proof}

It would seem natural to try and define a measure $\B_2$ of bias associated with $\S_2$. Considering equations
(\ref{eqn:fa2}), (\ref{eqn:ba2}) and (\ref{eqn:s2}), on would expect the following to define  $\B_2$:
\begin{equation}
1-\B_2(A)^2:=2\|A\|\,\|A'\|-\W(\sa)+\sqrt{\left[2\|A\|\,\|A'\|-\W(\sa)\right]^2-\left[2\S_0(A)-\S_1(A)\right]}.
\end{equation}
Using the notation introduced in the proof of theorem \ref{thm:s2}, we then have:
\begin{equation}
\S_2(A)=X-\sqrt{X^2-Y},\quad1-\B_2(A)=X+\sqrt{X^2-Y}.
\end{equation}
Mimicking the definition of equation (\ref{eqn:x}), we would expect the quantity $x$ there to be 
given by
\begin{equation}\begin{split}
\B_1(A)^2\equiv x^2&=\F_2(A)^2\B_2(A)^2=(1-X+\sqrt{X^2-Y})(1-X-\sqrt{X^2-Y})\\
&=(1-X)^2-(X^2-Y)=1-2X+Y.
\end{split}
\end{equation}
It is evident that $\B_1(A)^2\ge0$ is equivalent to $\B_2(A)^2\ge 0$, or $1-\B_2(A)^2\le 1$. However, it turns out that
these inequalities are not satisfied for all effects $A$. In fact, using the identity established in the proof of theorem
\ref{thm:s2}, 
\begin{equation}
Y=X-2\mu(\saap),
\end{equation}
we find:
\begin{equation*}\begin{split}
1-2X+Y&=1-X-2\mu(\saap)\\
&=-2(M-\tfrac 12)(\tfrac 12-m)
+\min_{\lambda\in\sa}(\lambda-\tfrac 12)^2+\max\left\{(M-\tfrac 12)^2,(m-\tfrac 12)^2\right\}\\
\end{split}
\end{equation*}
In the case where $M=1-m$ ($A$ unbiased) we find:
\[
1-2X+Y=-(M-\tfrac 12)^2+\min_{\lambda\in\sa}(\lambda-\tfrac 12)^2<0
\]
whenever $\sa\ne\{m,M\}$. It follows that the above expressions for $\B_1$ and $\B_2$ fail to give
bias measures.

We thus return to the equations (\ref{eqn:fa2}) and (\ref{eqn:ba2}) and try to find ways of writing these in a form applicable
to arbitrary Hilbert spaces. Observe that 
\[
a_0=\mu(\sa)=1-\mu(\sap),
\]
which we can take as a redefinition of $a_0$. There are at least two ways of rewriting the quantity $a_0^2-|\va|^2$:
\[
a_0^2-|\va|^2=M(\sa)m(\sa)=Mm\,.
\]
or 
\[
a_0^2-|\va|^2=a_0^2-\min_{\lambda\in\sa}(\lambda-a_0)^2=\max_{\lambda\in\sa}(2a_0\lambda-\lambda^2)=\|2\mu(\sa)A-A^2\|.
\]
and similarly
\[
(1-a_0)^2-|\va|^2=\|2\mu(\sa')A'-A'^2\|\,.
\]
In the first case we obtain:
\begin{equation}\begin{split}
\F_3(A)&:=\sqrt{M(\sa)m(\sa)}+\sqrt{M(\sap)m(\sap)}=\sqrt{Mm}+\sqrt{(1-M)(1-m)}\,;\\
\B_3(A)&:=\sqrt{M(\sa)m(\sa)}-\sqrt{M(\sap)m(\sap)}=\sqrt{Mm}-\sqrt{(1-M)(1-m)}\,.
\end{split}
\end{equation}
In the second case we have (noting that $\min_{\lambda\in\sap}(\lambda-a_0)^2=\min_{\lambda\in\sa}(\lambda-a_0)^2$):
\begin{equation}\begin{split}
\F_4(A)&:=\sqrt{a_0^2-\min_{\lambda\in\sa}(\lambda-a_0)^2}+\sqrt{(1-a_0)^2-\min_{\lambda\in\sa}(\lambda-a_0)^2}\,;\\
\B_4(A)&:=\sqrt{a_0^2-\min_{\lambda\in\sa}(\lambda-a_0)^2}-\sqrt{(1-a_0)^2-\min_{\lambda\in\sa}(\lambda-a_0)^2}\,.
\end{split}
\end{equation}
It is not hard to show that $\F_3(A)\in[0,1]$, $\F_4(A)\in[0,1]$, and $\B_3(A)\in[-1,1]$, $\B_4(A)\in[-1,1]$. In particular,
$\F_3(A)\le 1$ is a consequence of the Cauchy-Schwarz inequality:
\[
(\sqrt{M},\sqrt{1-M})\cdot(\sqrt{m},\sqrt{1-m})\le\sqrt{M+(1-M)}\sqrt{m+(1-m)}=1\,,
\]
and equality is reached exactly when $M=m$. Thus the condition $\F_3(A)=1$ holds if and only if $A=M\id$, that is,
$A$ is trivial.

The equation $\F_3(A)=0$ is equivalent to $m=0$, $M=1$. This does not require $A$ to be a nontrivial projection, 
so $\F_3$ fails to be an unsharpness measure. 

The equation $\F_4(A)=0$ is equivalent to $a_0=\min_{\lambda\in\sa}|\lambda-a_0|=1-a_0$. This, in turn, is equivalent
to $a_0=\frac 12=\min_{\lambda\in\sa}|\lambda-a_0|$, that is, $\sa=\{0,1\}$. Thus $A$ is a nontrivial projection.

However, $\F_4(A)=1$ is equivalent to $\min_{\lambda\in\sa}|\lambda-a_0|=0$, that is, $a_0\in\sa$. This does not require
$A$ to be trivial. Hence $\F_4$ fails to be an unsharpness measure.

The equation $\B_3(A)=0$ is equivalent to $Mm=(1-M)(1-m)$, that is, $M+m=1$, or $\mu(\sa)=\frac 12$. Thus (B2) is satisfied.
Further, $\B_3(A)=1$ holds if and only if $M=m=1$ or $A=\id$; likewise, $\B_3(A)=-1$ if and only if $M=m=0$, i.e., $A=\nul$. So (B3) holds.

Next, $\B_4(A)=0$ is equivalent to $a_0=1-a_0$, that is, $a_0=\frac 12$, so (B2) holds. Finally, $\B_4(A)=1$ ($\B_4(A)=-1$)
if and only if $a_0=1$ ($1-a_0=1$), and so $A=\id$ ($A=\nul$). Thus (B3) is fulfilled. 

It is possible to combine the virtues of $\F_3$ and $\F_4$ to obtain an unsharpness measure and an associated bias measure.
\begin{theorem}
The following defines an unsharpness measure on $\eh$:
\begin{equation}\begin{split}
\F_5(A)&:=\sqrt{\tfrac 12(a_0^2-\min_{\lambda\in\sa}(\lambda-a_0)^2)+\tfrac 12Mm}\\
&\qquad+\sqrt{\tfrac 12((1-a_0)^2-\min_{\lambda\in\sa}(\lambda-a_0)^2)+\tfrac 12(1-M)(1-m)}\,.
\end{split}
\end{equation}
Similarly, the following defines a bias measure on $\eh$:
\begin{equation}\begin{split}
\B_5(A)&:=\sqrt{\tfrac 12(a_0^2-\min_{\lambda\in\sa}(\lambda-a_0)^2)+\tfrac 12Mm}\\
&\qquad-\sqrt{\tfrac 12((1-a_0)^2-\min_{\lambda\in\sa}(\lambda-a_0)^2)+\tfrac 12(1-M)(1-m)}\,.
\end{split}
\end{equation}
Here the abbreviations $a_0=\mu(\sa)$, $M=\max(\sa)$, $m=\min(\sa)$ are used.
\end{theorem}
\begin{proof}
We note that 
\begin{equation}\label{eqn:ge1}\begin{split}
a_0^2-\min_{\lambda\in\sa}(\lambda-a_0)^2&=\tfrac 14(M+m)^2-\min_{\lambda\in\sa}(\lambda-a_0)^2\\
&=\tfrac 14(M-m)-\min_{\lambda\in\sa}(\lambda-a_0)^2+Mm\ge Mm\,,
\end{split}
\end{equation}
and similarly
\begin{equation}\label{eqn:ge2}
(1-a_0)^2-\min_{\lambda\in\sa}(\lambda-a_0)^2\ge(1-M)(1-m)\,.
\end{equation}
Note that equality holds in both cases if and only if $\sa=\{M,m\}$.
This entails that 
\begin{equation}\label{eqn:ge3}
\F_3(A)\le\F_5(A)\le\F_4(A)
\end{equation}
and $\F_5(A)\in[0,1]$, $\B_5(A)\in[-1,1]$. It is now a simple consequence of inequalities (\ref{eqn:ge1})-(\ref{eqn:ge3})
that $\F_5(A)=1$ if and only if $\F_3(A)=\F_4(A)=1$, and $\F_5(A)=0$ if and only if $\F_3(A)=\F_4(A)=0$. This ensures that 
$\F_5$ is in fact a sharpness measure. It is equally straightforward to verify that $\B_5$ is a bias measure.
\end{proof}

\section{Conclusion}

We have determined a variety of measures of the sharpness and bias of an effect. It was found that generalization
of such measures from $\C^2$ to arbitrary Hilbert spaces $\hi$ is not unambiguous in that there are different extensions.
Neither are such generalizations entirely trivial to construct: we encountered some suggestive candidates which nevertheless failed to possess the desired properties of sharpness or bias measures, indicating that the two-dimensional case may not yield all
relevant features of such measures.

A recently proven criterion for the coexistence of two qubit effects suggests that in general besides unsharpness, the 
bias of an effect is a significant quantity. This criterion involves certain measures of unsharpness and bias in a perfectly 
symmetric way, with closely related definitions, and we have found an extension of both measures to a general Hilbert 
space. It is an open question whether these or similar measures are relevant for the coexistence of effects in Hilbert spaces of dimension greater than two. It would be interesting to investigate the coexistence of more than two effects and of pairs of general POMs and to find out whether necessary and sufficient conditions can still be cast in the form of inequalities involving unsharpness and bias and possibly other quantities.

\vspace{6pt}
\noindent{\em Acknowledgement.} Part of this work was carried out during my visiting 
appointment at Perimeter Institute (2005-2007). Hospitality and support by PI are gratefully 
acknowledged.

\end{document}